\documentclass[11pt,hyper]{JHEP3}
\usepackage{txfonts,graphicx,verbatim}

\newcommand{\ds}{{\sf DarkSUSY}}

\title{Dark Matter Signals from Draco and Willman~1: Prospects for MAGIC~II and CTA} 

\author{Torsten Bringmann \\ Department of Physics, Stockholm
  University, AlbaNova, University Centre, S-106 91 Stockholm, Sweden
  \\ E-mail: \email{troms@physto.se}} 
\author{Michele Doro \\ Department of Physics G.~Galilei, University
  of Padova \& INFN, via Marzolo 8, 35131 Padova, Italy \\ E-mail:
  \email{michele.doro@pd.infn.it}} 
\author{Mattia Fornasa \\ Department of Physics G.~Galilei, University
  of Padova \& INFN, via Marzolo 8, 35131 Padova, Italy \\ Institut
  d'Astrophysique de Paris, boulevard Arago 98bis, 75014, Paris,
  France \\ E-mail: \email{mfornasa@pd.infn.it}} 

\preprint{\astroph{0809.2269}}

\abstract{The next generation of ground-based Imaging Air Cherenkov Telescopes 
will play an important role in indirect dark matter searches.
In this article, we consider two particularly promising 
candidate sources for dark matter annihilation signals, the nearby
dwarf galaxies Draco and Willman~1, and study the prospects of detecting such 
a signal for the soon-operating MAGIC~II telescope system as well as for 
the planned installation of CTA, taking special care of describing the 
experimental features that affect the detectional prospects.
For the first time in such studies, we fully take into account the
effect of internal 
bremsstrahlung, which has recently been shown to considerably enhance, in some
cases, the gamma-ray flux in the high energies domain where
Atmospheric Cherenkov Telescopes operate, thus leading to significantly harder 
annihilation spectra than traditionally considered. While the detection of the
spectral features introduced by internal bremsstrahlung would
constitute a smoking gun signature for dark matter annihilation, we
find that for most models the overall flux still remains at a level that will 
be challenging to detect, unless one adopts somewhat favorable 
descriptions of the smooth dark matter distribution in the
dwarfs.}


\begin{document}

\section{Introduction}
Even though basically nothing is known about its underlying nature,
compelling evidence has accumulated in recent years that the building
block of the observed structures in the universe is a new form of
non-baryonic, cold dark matter (DM), with the fraction of DM
to critical density being $\Omega_{\mbox{\tiny{DM}}}\sim 0.233 \pm 0.013$ 
according to the most recent estimates \cite{Komatsu:2008hk}. A class of
particularly well-motivated DM candidates arises in many
extensions of the standard model of particle physics in the form of
weakly interacting massive particles (WIMPs) that, thermally produced
in the early universe, automatically acquire a relic density matching
the observed DM abundance today
(for reviews, see \cite{Jungman:1995df,Bergstrom:2000pn,Bertone:2004pz,Taoso:2007qk}).    
In the following, we will restrict our analysis to the supersymmetric 
neutralino, which is maybe the best-motivated and the most widely studied DM
candidate.

\par
The search for DM is a vast field where different experimental
techniques can provide complementary information. At the Large Hadron
Collider (LHC), the detection of new supersymmetric particles through
missing energy signatures, would provide valuable information about
the composition of the neutralino and its role as a DM candidate 
(see, e.g. \cite{Battaglia:2004mp,Baltz:2006fm}). The
neutralino could also be observed through nuclear recoils in direct
detection experiments. A new generation of cryogenic detectors is
aiming to improve the already impressive recent limits reported by
Angle et al.~\cite{Angle:2007uj} and Ahmed et al.~\cite{Ahmed:2008eu}, thereby 
probing significant regions of the underlying parameter space. Finally, DM can
be searched for through \emph{indirect detection}, i.e.~by looking for  
primary and secondary products (mainly gamma rays, neutrinos, positrons and 
anti-protons) of DM annihilation. Our attention will be focused on the
detection of high-energy gamma rays.

In this field, Imaging Air
Cherenkov Telescopes (IACTs) play a leading role, exploring the
possibility of tracing Cherenkov photons from electromagnetic showers
produced by gamma rays impinging on the Earth's
atmosphere. Ground-based gamma ray astronomy is a very active
field that has matured rapidly in recent years. In the
near future, new installations with improved sensitivity will
investigate those regions in the sky where DM signatures can be
expected. In particular MAGIC~II, already in the commissioning phase,
will perform observations in the northern hemisphere, while a new
generation of  IACTs is currently in the design phase, the Cherenkov
Telescope Array (CTA), where an array of several tens of telescopes
will have improved angular and energy resolution, as well as a
very low energy threshold and a greatly increased sensitivity.

\par
Dwarf spheroidal galaxies (dSphs) are the faintest known 
astrophysical objects believed to be dominated by DM 
\cite{Gilmore:2007fy}.
So far, almost all detected dSphs  are located in the Local Group, a fact 
that is most likely related to the low luminosity of these objects, ranging 
from $330$~L$_\odot$ to $3\times10^7$~L$_\odot$ 
\cite{Mateo:1998wg,DaCosta:1999,Simon:2007dq,Geha:2008zr}. Typically, dSphs 
are satellite galaxies hosted by the halo of a larger and more massive galaxy 
that, in some cases, strongly influences their evolution: tidal 
stripping,  e.g., can deprive a dSph from stars in the outer  region 
(see, e.g., \cite{Munoz:2005be,McConnachie:2006nb}). While dSphs share 
many characteristics with stellar globular clusters --- a similar number of 
stars (from tens to a few thousand) and a  similar range in luminosity --- 
they usually have a larger size, a fact that implies the presence of a
DM halo of the order of  $10^6-10^8$~M$_{\odot}$. The resulting
mass--to--light ratios are typically larger than
10~M$_{\odot}/$L$_\odot$ and can even reach values up to
$10^3$~M$_{\odot}/$L$_\odot$, suggesting that these objects are
particularly interesting targets for indirect DM searches
(for previous studies, see e.g., \cite{Bergstrom:2005qk,Strigari:2007ma,Colafrancesco:2006he,SanchezConde:2007te,Strigari:2006rd}). 

\par
Let us mention in passing that the number of detected dSphs is around
one order of magnitude below what is expected from $N$-body simulations
of structure formation in $\Lambda$CDM cosmologies, which has become
known as the \emph{missing satellites problem} 
\cite{Klypin:1999uc,Moore:1999nt}. While many interpretations have
been suggested, including the effect of decaying DM 
(see, e.g., \cite{Cembranos:2005us,Kaplinghat:2005sy,Borzumati:2008zz} and references therein), the most natural explanation would be that the
missing dSphs have so far simply escaped detection. In fact,
it has been proposed that dSphs below a certain mass cannot
efficiently accrete baryons, which would explain the intrinsic
faintness of these objects \cite{Strigari:2007ma}. The claimed
discrepancy has recently also been mitigated considerably by the
discovery of a bunch of  new ultra-faint galaxies in the SDSS data 
\cite{Simon:2007dq}. 

\par
In this paper, we provide an update on the possible future  of the
indirect detection of DM with IACTs. In particular, we discuss the
prospects for the detection of DM in dSphs with MAGIC II  and make
tentative predictions in view of the expected characteristics of
CTA. We choose to focus on two dSphs, Draco and the recently discovered 
ultra--faint Willman~1. While the choice of Draco is supported by the large 
set of available data, Willman~1 will be discussed as a very promising target 
for future detection. 

\par 
A key ingredient in estimating the gamma ray flux for IACTs  is the 
computation of the energy spectrum for photons produced by neutralino
annihilations. In Ref.~\cite{Bringmann:2007nk}, it was recently shown 
that internal bremsstrahlung (IB) provides an important, hitherto neglected,
contribution to the spectrum. While the effect of IB is largely model
dependent, it appears in general as a pronounced ``bump'' at energies
close to the kinematic cutoff at the  neutralino mass. The importance of this 
effect is two--fold: first, the flux at high energies, where IACTs are most 
sensitive, is significantly increased; secondly, the introduction of spectral 
features allows an easier discrimination of a DM source from potential
astrophysical sources located in the vicinity, whose spectrum is
usually a featureless power law. This is the first time that IB is
fully included in studying prospects of DM detection for IACTs.

\par
The paper is organised as follows: in Section~\ref{sec:sources},  we will briefly review some of the main properties of Draco and Willman~1 and discuss how their DM profiles 
are constrained by observations. The following Section~\ref{sec:iact} is dedicated 
to a short  description of the IACT technique, with particular attention to 
the features that are relevant for DM observations. In Section~\ref{sec:flux}, we 
introduce a set of five benchmark neutralinos, representatives for the relevant
regions of the parameter space, and estimate the corresponding flux 
based on the technical specifications for MAGIC~II and CTA. 
Results are discussed in Section~\ref{sec:results}, and in Section~\ref{sec:conclusions} we
present our conclusions.

\section{Phenomenology of Draco and Willman~1}
\label{sec:sources}

\subsection{Draco}
Draco is located at a distance of  $80\pm7$~kpc from the Earth 
\cite{Mateo:1998wg} and is one of the best known and most often
studied  dSph. Discovered by Wilson~\cite{Wilson:1954}, the first
estimation of its DM content was performed only 30 years later by
Aaronson \cite{Aaronson:1983} from the analysis of the stellar
velocity dispersion. Nowadays, a large set of data is available for  
Draco
\cite{Shetrone:2000gz,Aparicio:2001cj,Piatek:2002ua,Segall:2006nc,Walker:2007ju}.  
Even though the formation of the DM halo cannot be traced back with
great precision, it can be inferred that the system is  
composed of very old, low--metallicity stars, without any significant
sign of star formation in the last 2~Gyrs. 
At its short
distance from the Milky Way (MW) center, Draco has likely been stripped off
its outermost stars in the past; today, however, no signs of tidal interactions are observed \cite{Segall:2006nc}. 
From a kinematical analysis of a sample of 200 stellar line-of-sight velocities (with radii that range from 50~pc to 1~kpc) one can infer the DM profile: The
result of the fit (assuming a Navarro--Frenk--White profile, hereafter
NFW \cite{Navarro:1996gj}) indicates a virial mass of the order of 
$10^9$~M$_\odot$ \cite{Walker:2007ju}, with a corresponding
mass--to--light ratio of M/L$\,\gtrsim200$~M$_\odot$/L$_\odot$ that
characterises Draco as highly DM dominated. 

\par
Many groups have analysed the available data for Draco and modelised its
DM profile
\cite{Colafrancesco:2006he,SanchezConde:2007te,Tyler:2002ux,Kazantzidis:2003hb,Lokas:2004sw,Mashchenko:2005bj}.
We are going to discuss here the two extreme cases of a \emph{cusp}
profile and a \emph{core} profile, which span the range of possible
configurations.\footnote{It should be
  underlined that core profiles are not compatible with current CDM
  simulations: recent results predict DM halos well-fitted by an
  Einasto profile without converging to a precise value for the inner
  slope~\cite{Springel:2008cc}. For comparison with previous studies, 
  however, we chose to still include core profiles in the analysis.} 
For the former, we take the NFW profile,

\begin{equation}
\rho_{\,\mbox{\tiny{NFW}}}\,(r) =\frac{\rho_s}{(r/r_s)\;(1+r/r_s)^2},
\end{equation}

\noindent
and for the latter we take a Burkert
\cite{Burkert:1995yz,Salucci:2000ps} profile,  
\begin{equation}
\rho_{\,\mbox{\tiny{Burkert}}}\,(r) =\frac{\rho_s}{(1+r/r_s)\;\left(1+(r/r_s)^2\right)}.
\end{equation}

For the scale radius $r_s$ and the scale density $\rho_s$, we use the 
values summarised in Table~\ref{tab:profiles} (taken from
Ref.~\cite{Mashchenko:2005bj}).  
Note that both profiles produce good fits to the velocity dispersion
profiles, down to the pc scales where the innermost stars are
observed. No direct observational information is available for
distances even closer to the centre, where the two profiles differ
significantly.  

\par
Possible DM annihilation signals from Draco have already been
searched for in the past: after the CACTUS experiment had claimed the
detection of an excess of $\sim7000$ high-energy photons in
only $7$~hours \cite{Chertok:2006yp}, almost all IACTs tried to
reproduce the result, but the claim was not confirmed
\cite{Driscoll:2007ea,Wood:2008hx}. MAGIC observed Draco for
$7.8$~hours during 2007 above  $140$~GeV \cite{Albert:2007xg}. Within
a 2$\sigma$ confidence limit, the collaboration reported an upper limit
for the integral flux  of $\Phi^{u.l.}(E>140$~GeV$)<1.1\times
10^{-11}$~ph cm$^{-2}$ s$^{-1}$.

\subsection{Willman 1}
Willman~1 (SDSS J1049+5103) is a very peculiar object, located
at a distance of $38\pm7$~kpc from the Earth in the constellation of 
Ursa  Major. Discovered in 2004 by B.~Willman
\cite{Willman:2005cd,Willman:2006mv}, using  
data from the Sloan Digital Sky Survey \cite{York:2000gk}, it was
then further observed with Keck/DEIMOS \cite{Martin:2007ic} and more
recently by Siegel et al.~\cite{Siegel:2008tz}.  With an absolute magnitude of 
M$_V\sim-2.5$ and a half--light radius (i.e. the radius of a cylinder, pointing
to the Earth, that encloses half of the luminosity of the object) of
$21\pm7$~pc, it looks very similar to a globular cluster, even if its
narrow distribution of stellar velocities and the large spread in
stellar metallicities suggests that it is indeed the smallest dSph ever observed.
The object may show evidence for tidal disruption from its tri-axial stellar
distribution \cite{Willman:2006mv}. On the other hand, the difference
in the stellar luminosity function between the  central and outermost
stars reveals a strong mass segregation. With a luminosity of
$855$~L$_\odot$, and a  mass of the order of $5\times 10^5~$M$_\odot$   
\cite{Martin:2007ic}, Willman~1 could feature a  mass--to--light 
ratio in the range $500-700$~M$_\odot/$L$_\odot$, or even more, making  
it one of the most DM dominated objects in the Universe 
\cite{Strigari:2007at}.  

\par
The small number of stars that belong to this dSph hinders, however,
an accurate determination of the DM density profile. Following 
Strigari et al.~\cite{Strigari:2007at}, we parametrise its DM halo with an NFW profile, 
as  specified in Table~\ref{tab:profiles}, although these parameters
are subject to somewhat larger uncertainties than in the case of Draco.  

\TABULAR[hbt!]{lccc}{
\hline
 & Draco--Burkert & Draco--NFW & Willman~1--NFW \\
\hline
$\;r_s\; (\mbox{kpc}) $ & 0.35 & 0.50 & 0.18 \\
$\;\rho_s\; (\mbox{M}_\odot/\mbox{kpc}^3)$  & $ 3.6\times 10^8 $ & $ 1.3\times10^8 $ & $ 4.0\times10^8 $\\
$\;\theta_{90}\; (^\circ) $ & 0.52$^\circ$ & 0.35$^\circ$ & 0.20$^\circ$ \\
$\;D\; (\mbox{kpc})$ & \multicolumn{2}{c}{$80\pm7$} & $38\pm7$ \\
\hline
}
{\label{tab:profiles} Scale radius (kpc) and scale density 
(M$_\odot$/kpc$^3$) that appear in the DM density profiles. The NFW and the 
Burkert profiles in the case of Draco represent models N3 and B2, 
respectively, from Ref.~\cite{Mashchenko:2005bj}. In the case of Willman~1, 
the NFW fit is taken from Ref.~\cite{Strigari:2007at}. In addition, the 
semi-aperture of the solid angle corresponding to 90\% of emission 
$\theta_{90} $ ($^\circ$) is reported together with the target distance $D$ 
(kpc).}

\section{Observations with IACTs: MAGIC~II and CTA}
\label{sec:iact}

When entering into the atmosphere, cosmic gamma rays
(as well as the many orders of magnitude more frequent
background charged cosmic rays) quickly lose energy through interactions
with the nuclei of atmospheric 
molecules, dominantly by pair production of electrons and positrons. 
Bremsstrahlung photons radiated by these highly energetic electrons and 
positrons in turn lead to the production of secondary electron-positron pairs, 
thus triggering the subsequent development of a particle shower.
When the electron--positron energy falls below 
$E_c\approx83$ MeV in the atmosphere, the dominant mechanism of energy
loss becomes ionization and the shower
rapidly dies off. This takes place at an altitude of $8-12$ km,
depending on the energy of the primary gamma ray.  
This cascade of highly relativistic particles causes a 
flash of UV-blue Cherenkov light, with the greatest emission coming
from the shower maximum (i.e.~where the number of free electrons and
positrons is maximal), lasting  a few nanoseconds and  propagating   
in a cone with an opening angle of $\sim1^{\circ}$, slightly depending
on the primary energy. The resulting circle of 
projected light at 2000 m asl (the MAGIC telescope altitude) has a radius of 
about $\sim120$~m. If a telescope is located inside this
Cherenkov-light pool, the light can be reflected from the collecting
mirrors and focused onto a multi-pixel recording camera. An image reconstruction
algorithm~\cite{Hillas:1985} then allows the recovery of the energy and
direction of the primary particle, and determines whether it was more
likely a hadron 
or a photon. In this way, it is possible to reject up to
99\% of the background, constituted mainly by sub--showers generated
by charged cosmic ray particles, by muons  and by the night sky
background light. 

\par
This technique was pioneered by the Whipple telescope \cite{Kidea:2007},
with several successors currently operating, including
MAGIC, HESS, CANGAROO-III and VERITAS\footnote{See, respectively,
\href{http://wwwmagic.mppmu.mpg.de}{http://wwwmagic.mppmu.mpg.de},
\href{http://www.mpi-hd.mpg.de/htm/HESS}{http://www.mpi-hd.mpg.de/htm/HESS},
\href{http://icrhp9.icrr.u-tokyo.ac.jp/}{http://icrhp9.icrr.u-tokyo.ac.jp/}
and 
\href{http://veritas.sao.arizona.edu/}{http://veritas.sao.arizona.edu/}}.    
In this paper, we will address observational 
prospects for the upcoming MAGIC~II telescope system and for the future
generation of IACTs,  focusing on the case of CTA.  MAGIC II is a
stereoscopic system of telescopes, composed of MAGIC and  
a second telescope currently under commissioning on the island of La Palma,
which will start operation in 2009.
The stereoscopic view of two telescopes
(pioneered by HEGRA \cite{Pulhofer:2003}), together with the
improved technical characteristics  
of the second detector, will allow a general improvement in the overall 
performance of the experiment, in particular in terms of energy and angular
resolution, as well as energy threshold.
The performance of the MAGIC~II array was simulated with Monte
Carlo (MC) tools by Carmona et al.~\cite{Carmona:2007rs}. CTA, on the other 
hand, is the result of an effort for a next generation Cherenkov observatory 
with increased capabilities: normally, one single telescope can 
cover 1.5--2 orders of magnitude in energy range. With the combined
use of many telescopes of 2--3 different sizes, CTA should be able to
extend the energy range to almost 4 orders of magnitude, from roughly 
$\sim30$~GeV to $\sim100$~TeV. The experiment is still in the early design 
phase and the final layout of the array is thus far from defined yet; the
performance, therefore, is still subject to changes. For this  
paper, we will refer mainly to the work of Bernl\"oher et
al.~\cite{Bernloher:2007} and 
several private communications within the collaboration. The CTA
prototype construction could start in 2010, at least for
some of the main components, and the final installation is foreseen in
2012--15.

\par   
The performance of an IACT in terms of its prospects to detect a DM 
annihilation signal can generally be characterised by a small number
of basic parameters, which are described in the following (see also 
Table~\ref{tab:IACT} for a summary of the characteristics for MAGIC~II and CTA):

\begin{itemize}
\item {\it Energy threshold:} The energy threshold of an IACT can take
  slightly different values according to the definition. Hereafter we
  consider it to be the peak of the reconstructed MC energy
  distribution (other definitions being analysis threshold, trigger
  threshold, etc.). This value depends mainly on the reflector area of
  the telescope: A larger mirror area allows, in particular, to
  collect more photons from the showers and thus increases the chance
  of discrimination against the night sky background light, in
  particular for low energy showers. The use of a stereoscopic system
  also plays an important role because it enhances the gamma/hadron 
  ($g/h$) discrimination power  which is weaker at low energy. The energy
  threshold changes with the zenith angle of observation, and sources
  culminating high in the sky are preferred. Reaching a low energy
  threshold is an important feature for DM studies with IACTs, both
  because of the increased number of photons and because of the
  enhanced possibility to observe the spectral cutoff even for
  low-mass neutralinos. Making use of stereoscopic observations,
  MAGIC~II will have an energy threshold of 60-70 GeV
  \cite{Carmona:2007rs},  with possible extension to even lower
  energies with improved analysis techniques and new trigger systems
  currently under development \cite{Schweizer:2007}. This value will
  be further lowered to at least 30~GeV for CTA. The telescope
  acceptance for gamma rays around 30~GeV starts to decrease rapidly,
  but a very strong gamma ray signal could probably even be detected
  at energies as low as about 10~GeV. 

\item {\it Energy resolution:} The true energy of the primary gamma
  ray $E'$ is reconstructed on the basis of a comparing analysis
  between the shower image parameters and MC events. The probability
  to assign, after the analysis, an energy $E$ to the primary gamma
  ray can be approximated by

  \begin{equation}\label{eq:E_res}
    R_\epsilon(E-E')\approx\frac{1}{\sqrt{2\pi}E}\cdot \exp\left(-\frac{(E-E')^2}{2\epsilon^2E^2}\right)\,.
  \end{equation}

  \noindent
  Typical values for the energy resolution $\epsilon$ are of the order
  of 10-30\% for IACTs, depending on the energy. The reason for such
  large uncertainties is the combined effect of many sources of
  uncertainties (for a more detailed discussion, see 
  Ref.~\cite{Albert:2007xg}). The energy resolution is an important 
  parameter when observing spectral features as bumps and cutoffs that can
  provide clear signatures for a DM signal. MAGIC II will have an energy 
  resolution of 15\% above 300 GeV (up to 20\% at 70 GeV); for the CTA,
  this situation could radically improve. Finally, let us note that further
  systematic errors might hide in the absolute energy calibration; the
  recent MAGIC observation of a clear cutoff in the Crab Nebula
  spectrum\footnote{The Crab Nebula is a supernova remnant that is
  conventionally taken as reference source for cross--calibrations in
  gamma ray astronomy due to its very stable and intense flux.}, when
  compared to a corresponding future observation by Fermi-LAT
  \cite{Gehrels:1999ri}, may allow for the first robust calibration
  of gamma ray energies \cite{Biland:2008}.

\item {\it Angular resolution:} The reconstruction of the direction  
  of a primary gamma ray is performed through image analysis. As a 
  result, a gamma ray coming from a direction $\psi^\prime$ will be 
  reconstructed to a direction  $\psi$ in the sky with a probability
  distribution that can be fitted to a Gaussian function:

  \begin{equation}\label{eq:ang_res}
    B_{\vartheta_r}(\psi'-\psi)=\frac{1}{2\pi\vartheta^2_r}\cdot 
    \exp\left(-\frac{(\psi'-\psi)^2}{2\,\vartheta^2_r}\right).
  \end{equation}

  \noindent
  The standard deviation $\vartheta_r$ of the Gaussian corresponds to
  the telescope angular resolution, also called Point Spread Function
  (PSF). As a consequence, any source will appear somewhat
  blurred. The stereoscopic system  exploited in MAGIC~II will improve
  the PSF, allowing values as low as $0.05^{\circ}$, while for CTA we
  expect an even smaller PSF. It is hard to predict an exact value
  given the current lack of knowledge of the CTA design, but a
  realistic value that we use for this study is $0.02^{\circ}$
  (see also \cite{Hofmann:2006wf}). For extended sources, as in the
  case of dSphs, the PSF plays an important role in the reconstruction
  of the DM density profile, as discussed in the next section.  

\item {\it Flux sensitivity:} The sensitivity of an IACT is usually
  defined as {the minimum flux for a $5\,\sigma$ detection over the
  background, after 50 hours of observation time and based on at least
  10 collected photons}.  For operating experiments, the sensitivity
  can be computed by using real data and following Eq. 17 of
  \cite{Li:1983fv}, while for planned experiments the sensitivity has
  to be estimated on the basis of MC simulations and is therefore
  subject to larger uncertainties. The procedure is as follows: a full
  data analysis is performed on two samples of MC simulations, one for
  gamma ray  events and one for background events (basically protons
  and helium), during which a number of parameters (``cuts'') is
  optimised to maximise the analysis quality factor
  $Q=\epsilon_\gamma/\sqrt{\epsilon_h}$, i.e. the ratio between the
  efficiency for gamma rays and the square root of the hadron
  efficiency (``efficiency'' refering here to the ratio between the
  number of events  passing the analysis cuts and the number of events
  at MC, input, level). After the optimisation, one can estimate the
  number of hadrons $N_h(>E)$ above some energy $E$. Given the
  Poissonian distribution of events, a $5\,\sigma$ detection is
  obtained whenever the number of gamma rays  detected is larger than
  $5\,\sqrt{N_h\,(>E)}$. The integrated sensitivity above  $E$ is thus
  given by:  

  \begin{equation}\label{eq:sens2}
    \Phi^{min}(>E)=\frac{5\sqrt{N_h\,(>E)}}{A\cdot t_{50}}
    \frac{1}{\epsilon_\gamma}\,,
  \end{equation}

  \noindent  
  where $A$ is the MC gamma ray simulation area and $t_{50}$ is the time 
  interval corresponding to 50 hours.

  $N_h(>E)$ and $\epsilon_\gamma$, and thus the sensitivity, are
  usually determined assuming a featureless power law spectrum of MC
  gamma ray events of index $-2.6$. This corresponds approximately to
  the spectrum of the Crab. For this reason, the sensitivity is often
  also expressed in terms of ``Crab'' units (C.U). In the case of the
  benchmark neutralinos under study (see next section), the gamma ray
  spectra are usually harder than that of the Crab and no longer
  featureless; it is therefore natural to ask how much this would
  change the sensitivity.  

  To address this question, let us note that the sensitivity mainly
  depends on the $g/h$ discrimination power. The $g/h$ separation,   
  however, is very efficient at intermediate and large energies, where
  the shower parameters are firmly distinguishable between hadronic
  and gamma events. At energies below $\sim30$~GeV, on the other hand,
  the differences are more subtle and the sensitivity \emph{is}
  affected. Hence we expect that the differential sensitivity does
  not depend too strongly on the spectrum of the source, unless in the
  case of rather low energies. An exact treatment of this effect would
  require dedicated studies with MC simulations, which is beyond the
  aim of this work. Based on a preliminary MC analysis, however, 
  we generally expect that the sensitivity at a given energy will not
  change by more than a factor of two compared to that defined for the
  Crab. A full analysis of this effect is left for future work.
\end{itemize}

\TABULAR[hbt!]{l|ccc} 
{ \hline
  & MAGIC & MAGIC II & CTA$^*$ \\
  \hline
  E$_0$ (GeV) & 100  & 70  & 30  \\
  $\epsilon$ & 30-20\% & 20-10\% & 10\% \\
  $\vartheta_r\;(^\circ) $ & $ 0.10^{\circ} $ & $ 0.05^{\circ} $ & $ 0.02^{\circ} $ \\
  S$(>E_0)\;(\mbox{cm}^{-2}$ $\mbox{s}^{-1}) $ & $5\times 10^{-11}$ & $ 1.4\times 10^{-11}$ & $1.5\times 10^{-11}$ \\
  \hline } {\label{tab:IACT} Comparison of the performance of the
  MAGIC, MAGIC II and CTA$^*$ telescopes. E$_0$ (GeV) is the energy
  threshold, $\epsilon$ the energy resolution and $\vartheta_r
  (^\circ)$ the angular resolution. The sensitivity S$(>E_0)$
  (cm$^{-2}$ s$^{-1}$) is given for a Crab-like spectrum above the
  energy threshold. \newline $^*$For CTA, the numbers have to be taken
  as placeholders because the telescope design is not yet fixed.}

\section{Flux estimation}
\label{sec:flux}
The gamma ray flux from DM annihilation can be factorised into two
different contributions: 

\begin{equation}
\frac{d\,\Phi}{dE} = J(\psi) \cdot \frac{d\,\Phi^{PP}}{dE}\,.
\label{eq:flux}
\end{equation}

\noindent
The term $J(\psi)$, also called the {\it astrophysical factor}, 
depends on the DM morphology at the emission region. 
The {\it particle physics factor} $d\,\Phi^{PP}/dE$, on the other hand, depends
on the microscopic properties of the DM candidate, in particular its mass
and cross section, as well as the annihilation modes and branching
ratios, that define the gamma ray spectrum. 
For a given telescope energy threshold $E_0$, the integral flux is thus 
\mbox{$\Phi(>E_0) = J\cdot\Phi^{PP}(>E_0)$}.  

The ability to reconstruct the two factors depends on the experimental
performance of the IACT; in the following, we will estimate the expected flux 
for Draco and Willman~1.

\subsection{Astrophysical Factor}\label{subsec:astro}
The astrophysical factor depends on the source distance and geometry (as well 
as the PSF of the telescope), but for a given DM profile it does not depend  
on the particular DM candidate. As a consequence, the discussion here remains 
valid for any generic WIMP candidate. Pointing the telescope towards a 
direction $\psi$ in the sky, and taking into account its finite angular
resolution, the astrophysical factor is given by: 

\begin{equation}\label{eq:jpsi_2}
J(\psi) = \frac{1}{4\pi}
\int d\Omega\int d\lambda~\left[\rho^2(r(\lambda,\psi))
  \cdot B_{\vartheta_r}(\theta)\right]\,,
\end{equation}

\noindent
where the angular integration $d\Omega= d\varphi\,d(\cos\theta)$
extends over a cone centered around $\psi$, with an opening angle a
few times the PSF $\vartheta_r$. The integration over $\lambda$ is
along the line-of-sight, in the direction $\psi$, such that $
r=\sqrt{\lambda^2 + D^2 - 2 D  \lambda\cos(\Psi)} $, where $ D $ is
the distance of the source from the Sun and
$\cos(\Psi)\equiv\cos(\theta)\cos(\psi)-\cos(\varphi)\sin(\theta)\sin(\psi)$.
Defined as above, $J(\psi)$ is conventionally expressed in units of
M$_\odot^2\,$kpc$^{-5}\,$sr$^{-1}$ or
GeV$^2\,$cm$^{-5}\,$sr$^{-1}$. In order to translate it to the 
dimensionless quantity $J(\psi)$ as defined in Ref.~\cite{Bergstrom:1997fj}, 
one simply has to multiply it by   
$5.32\times10^{-21}\,\mathrm{GeV}^{-2}\mathrm{cm}^5\mathrm{sr}$
($=2.37\times10^{-14}\,\mathrm{M_\odot}^{-2}\mathrm{kpc}^5\mathrm{sr}$).  

Integrating Eq.~\ref{eq:jpsi_2} over the full angular extension of
the source gives:

\begin{equation}
\label{flux_point}
\widetilde{J}\equiv\int d\Omega_\psi\,J(\psi)\simeq\frac{1}{4\pi D^2}\int dV\;\rho^2(r)\,,
\end{equation}

\noindent
where the second integral is over the \emph{spatial} extent of the
source. Note that this expression is now given in units
M$_\odot^2\,$kpc$^{-5}$ (or GeV$^2\,$cm$^{-5}$) and no longer depends
on the telescope PSF.  

\FIGURE[htb!]{
\caption{\label{fig:profile} The $J(\psi)$ factor in
  the case of Draco (upper plot) and Willman~1 (lower plot). Core
  profiles are shown in blue, cusp profiles in red. Thick solid (dashed) lines
  represent profiles smeared with the MAGIC~II (CTA) angular
  resolution. Thin solid lines represent the profiles without smearing.
  The upper right panel  in each figure shows a zoom-in of the region
  close to the center. For comparison, we also show the profiles for a hypothetical, infinite angular resolution.} 
\includegraphics[width=0.75\linewidth]{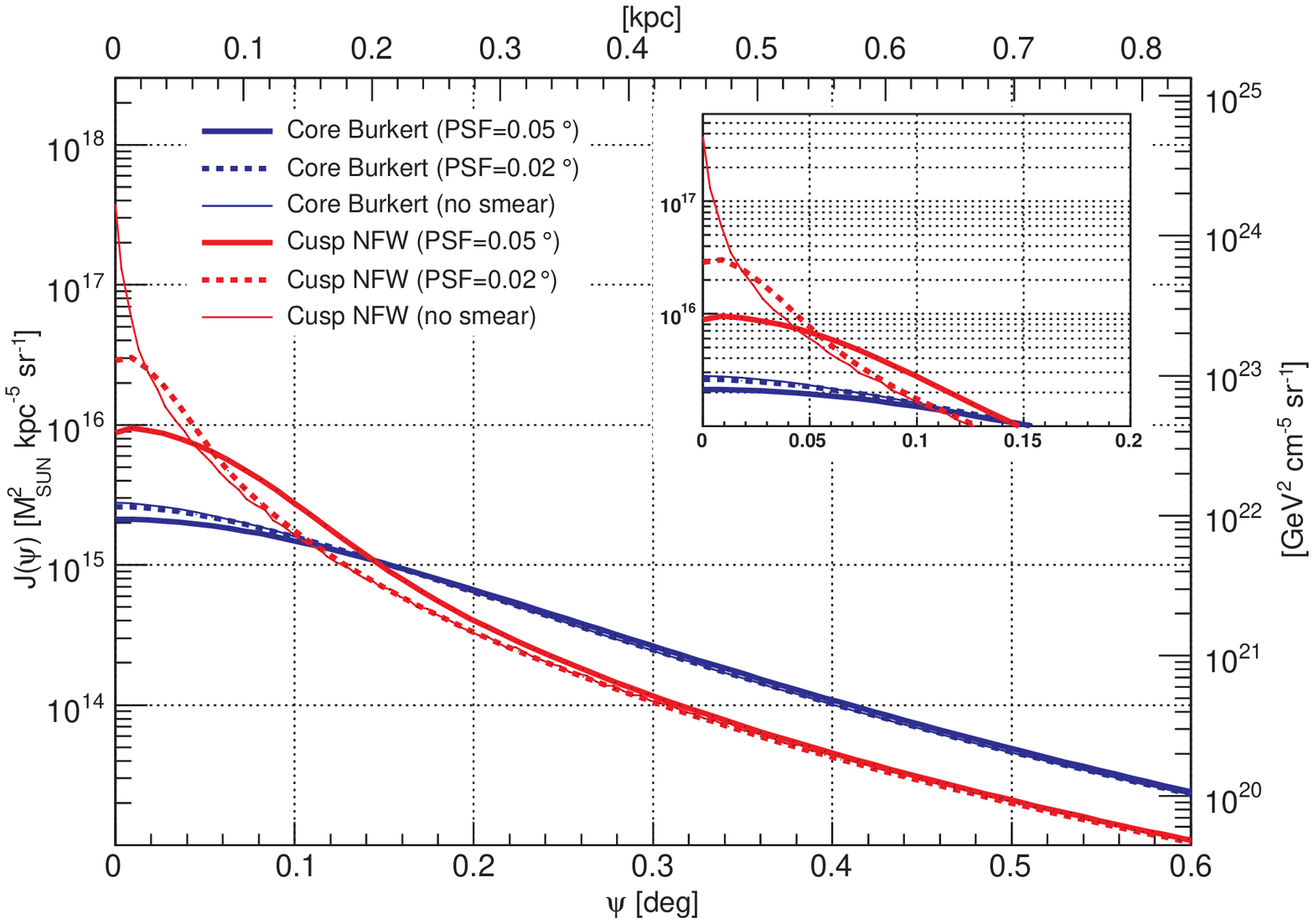}
\includegraphics[width=0.75\linewidth]{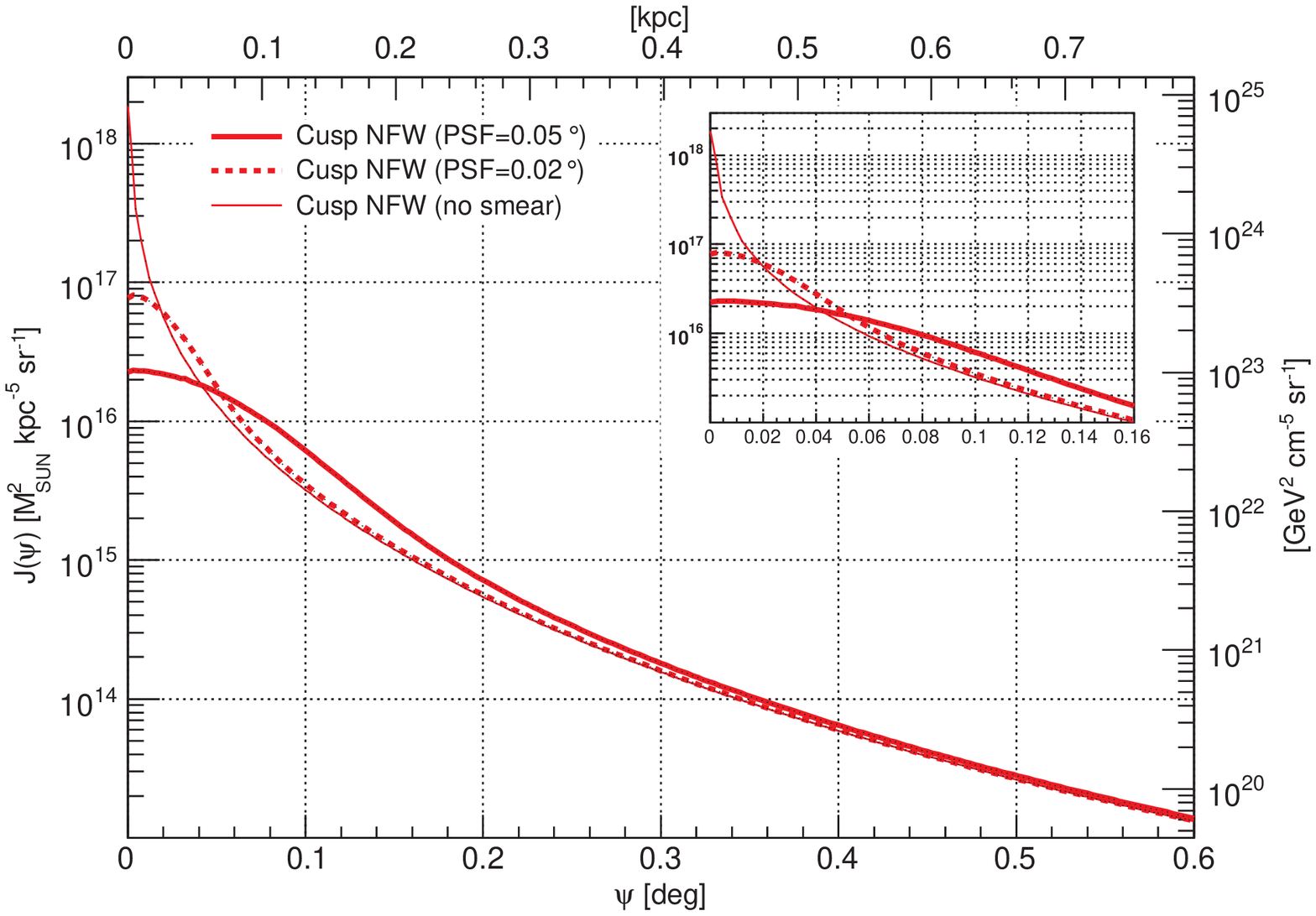}}

Using the DM profile parameters of Table~\ref{tab:profiles}, we now
show in Figure~\ref{fig:profile} the quantity $J(\psi)$ for Draco and Willman~1. 
While the two sources, from this plot, appear similar in 
terms of their angular size $\psi$, we recall that their spatial
extension is quite different: By comparing, e.g., their respective
scale radii for an NFW profile, one sees that Willman~1 ($r_s=0.18$~kpc) 
is considerably smaller than Draco ($r_s=0.50$~kpc).  
In the case of Draco, the cusp and core profiles are almost identical
(up to an overall normalisation factor of $\sim2$) for angular
distances larger than about $\psi\sim0.3^\circ$, below which the cusp
profile starts to increase more rapidly. At the center, the two profiles 
differ by around one order of magnitude, the difference increasing with 
decreasing PSF. Whenever an extended emission would be observed, one could 
thus in principle be able to discriminate between different profiles by  
comparing the flux at different distances from the center. As it becomes 
obvious from Figure~\ref{fig:profile}, and as already stressed by 
Sanchez--Conde et al.~\cite{SanchezConde:2007te}, the angular resolution of the telescope 
would play an important role in this case. Taking into account the full 
range of profiles consistent with the observational data 
\cite{Strigari:2007at}, we note that the astrophysical factor for dSphs is  
far better constrained than for, e.g., the galactic center, where the 
uncertainty in the inner part spans several orders of magnitude 
\cite{Fornengo:2004kj}. 

Given a telescope PSF of the order of $0.1^\circ$, and the expected feebleness 
of the signal, however, the capability of reconstructing the morphology of 
extended sources is very limited. This is particularly true in the case of 
non-stereo IACTs where the shower direction is reconstructed with less 
precision. 
Even when making the rather optimistic assumption that a signal could be 
discriminated against the background out to a distance where the annihilation 
flux is a factor of 3 less than from the direction towards the center, the 
source would appear at a size of only roughly twice the PSF for both
Draco and Willman~1, in the case of a cuspy profile. For a core profile, the 
same measure would indicate an apparent extension out to $\lesssim0.2^\circ$,  
still well contained in a normal IACT camera ($\sim3^\circ$ aperture). As we 
will see, the expected annihilation fluxes are rather low and we find it 
therefore premature to discuss in depth the  possibilities to distinguish 
between different profiles in the way indicated above; rather, we will in the 
following focus on the total, i.e. integrated, flux.

\TABULAR[htb!]{cccc}{
\hline
$ \widetilde{J}$ & Draco--Burkert & Draco--NFW & Willman~1--NFW \\
\hline (GeV$^2$/cm$^5$)  
 & $ \;3.84 \times 10^{17}\; $ 
 & $ \;4.71 \times 10^{17}\; $ 
 & $ \;9.55 \times 10^{17}\; $ \\
 (M$_\odot$/kpc$^5$)
 & $  8.63  \times 10^{10}$ 
 & $  1.06 \times 10^{11} $ 
 & $  2.15 \times 10^{11} $ \\
\hline
}
{\label{tab:jpsi} Comparison of the integrated astrophysical factor 
$\widetilde{J}$ for Draco and Willman~1, for the profiles specified 
in \ref{tab:profiles}.}

Table~\ref{tab:jpsi} reports the calculation of the \emph{integrated} 
astrophysical factor $\widetilde{J}$ for the two dSphs studied here. For 
Willman~1, the uncertainty in the DM profile translates into a 95\% 
confidence interval of about 
\mbox{$8\times10^{17}\,\mathrm{GeV}^2/\mathrm{cm}^5\lesssim\widetilde{J}
\lesssim4\times10^{19}\,\mathrm{GeV}^2/\mathrm{cm}^5$} 
\cite{Strigari:2007at}. In the case of Draco, the astrophysical factor 
lies in the range 
\mbox{$10^{17}\,\mathrm{GeV}^2/\mathrm{cm}^5\lesssim\widetilde{J} \lesssim2\times10^{18}\,\mathrm{GeV}^2/\mathrm{cm}^{5}$} 
\cite{Strigari:2006rd}.   
Again, these astrophysical uncertainties are rather small when compared to 
other potential sources of DM annihilation signals -- but one should keep in
mind that our choices of DM profiles are actually quite conservative: taking 
into account the above quoted range of possible values for $\widetilde{J}$ 
that are consistent with current observations of velocity dispersions in the 
dwarfs, one could thus win a factor of up to about 4 (in the case of Draco) 
or 40 (in the case of Willman~1) in the annihilation flux; we will get back to this in Section \ref{sec:results}.

\subsection{Particle Physics factor and Benchmarks}
\label{subsec:phipp}

\TABLE[htb!]{
\begin{tabular}{ccccccccc}
\hline\hline
      BM & $m_{1/2}$ & $m_0$ & $\tan\beta$ & $A_0$ & s($\mu$) & $m_{\chi}$ & $\Omega_{\chi} h^2$ & $\sigma v|_{v=0}$\\
\hline
$I'$  & 350  & 181   & 35.0 & 0    & $+$ & 141 & 0.12 & $3.6\cdot10^{-27}$ \\
$J'$  & 750  & 299   & 35.0 & 0    & $+$ & 316 & 0.11 & $3.2\cdot10^{-28}$ \\
$K'$  & 1300 & 1001  & 46.0 & 0    & $-$ & 565 & 0.09 & $2.6\cdot10^{-26}$ \\
$F^*$ & 7792 & 22100 & 24.0 & 17.7 & $+$ & 1926& 0.11 & $2.6\cdot10^{-27}$ \\
$J^*$ & 576  & 108   & 3.8  & 28.3 & $+$ & 233 & 0.08 & $9.2\cdot10^{-29}$ \\
\hline\hline
\end{tabular}
\newline\linebreak
\begin{tabular}{cccccc}
\hline\hline
BM  &  $r$ & $2\cdot\sigma v|_{\gamma\gamma}$ & $\sigma v|_{Z\gamma}$& $\Phi^{PP}_{>70~GeV} $ & $\Phi^{PP}_{>30~GeV}$ \\
\hline
$I'$  & 4     & $7.9\cdot10^{-30}$ & $8.5\cdot10^{-31}$ & $1.6\cdot10^{-33}$ & $9.9\cdot10^{-33}$\\
$J'$  & 34    & $3.4\cdot10^{-30}$ & $4.1\cdot10^{-31}$ & $2.2\cdot10^{-34}$ & $1.1\cdot10^{-33}$\\
$K'$  & $<$0.1& $1.8\cdot10^{-31}$ & $2.2\cdot10^{-32}$ & $1.5\cdot10^{-32}$ & $7.5\cdot10^{-32}$\\
$F^*$ & 11    & $5.8\cdot10^{-30}$ & $1.6\cdot10^{-29}$ & $9.6\cdot10^{-34}$ & $2.4\cdot10^{-33}$\\
$J^*$ & 2300  & $5.5\cdot10^{-30}$ & $1.8\cdot10^{-30}$ & $7.9\cdot10^{-34}$ & $8.9\cdot10^{-34}$\\
\hline\hline
\end{tabular}
\caption{\label{tab:bm} Parameters defining the benchmark models and some 
relevant quantities related to the annihilation spectrum. $m_{1/2}$ and $m_0$
(expressed in GeV) are the uniform masses of gauginos and scalars,
respectively. $\tan\beta$ is the ratio between the vacuum
expectation values of the two Higgs bosons. $A_0$ (in GeV) is the
coefficient of the trilinear scalar term and $\mu$ is the coefficient
of the mass term in the Higgs potential. $m_\chi$ (in GeV) is the
neutralino mass, $\Omega_\chi h^2$ its  relic density and
$\sigma v|_{v=0}$ (expressed in cm$^3$~s$^{-1}$) its
annihilation rate today. $r$ is the ratio of IB photons over
secondary photons (above $0.6\,m_\chi$) and $\sigma
v|_{\gamma\gamma}$, $\sigma v|_{Z\gamma}$ are the annihilation rates
for the $\gamma$ lines. $\Phi^{PP}$ (expressed in cm$^3$~s$^{-1}$~GeV$^{-2}$), 
finally, is defined in Eq.~\ref{eq:particle_physics} and given for MAGIC~II 
($E_0=70$ GeV) and CTA ($E_0=30$ GeV) energy thresholds, respectively.}
}

The particle physics factor in Eq.~\ref{eq:flux} is given by: 

\begin{equation}
\frac{d\Phi^{PP}}{dE}=\frac{\sigma v}{2m_\chi^2} \cdot 
\sum_i B^i \int dE'\frac{dN_{\gamma}^i (E')}{dE'}  R_\epsilon(E-E')\,,
\label{eq:particle_physics}
\end{equation}

\noindent
where $\sigma v$ is the total annihilation rate of the
DM particles, $m_\chi$ the mass of the DM particle, $B_i$ the branching ratio 
into channel $i$ and $dN_{\gamma}^i/dE$ the corresponding differential 
number of photons per (total) annihilation. The integration over
$R_\epsilon(E-E')$ (see Eq.~\ref{eq:E_res}), takes into account  
the finite energy resolution of the instrument. The \emph{total} number of 
photons above some energy $E_0$ of course no longer depends on the energy 
resolution (as long as $1-E_0/M_\chi\gg \epsilon$) and is given by:

\begin{equation}
N_\gamma(>E_0) \simeq  \sum_iB^i\int_{E_0}^{m_\chi}\frac{dN_{\gamma}^i
  (E)}{dE}\;dE \,.
\label{eq:particle2}
\end{equation}

Three different contributions to the spectrum can, in general, be
distinguished: first of all  \emph{monochromatic} $\gamma$
\emph{lines}, where photons are primary products of neutralino
annihilation through the reactions $\chi\chi\rightarrow\gamma\gamma$
and $\chi\chi\rightarrow\gamma Z^0$ \cite{Bergstrom:1997fj}.
These processes are very model-dependent; while providing a striking
experimental signature, they are usually subdominant 
(for a recent analysis, see \cite{Bringmann:2007nk}). As a further
contribution, \emph{secondary photons} are produced in  the hadronization and
further decay of the primary annihilation products, mainly through the
decay of neutral pions, resulting in a featureless spectrum with a
power law like behaviour at small photon energies and a rather soft
cutoff at $m_\chi$. These contributions, which always dominate the
total spectrum at low energies, are highly model-independent and
have a spectral shape that is almost indistinguishable for the various 
possible annihilation channels 
(see, e.g., \cite{Bertone:2004pz,Fornengo:2004kj}). 

For charged annihilation products, a third contribution has to be
included; \emph{internal brems\-strah\-lung}, where an additional photon
appears in the final state. As has been pointed out recently
\cite{Bringmann:2007nk}, these photons generically dominate the total
spectrum at energies close to the kinematical cutoff at $m_\chi$. Moreover, 
they add pronounced signatures to the spectrum that would
allow for a clear identification of the DM origin of an observed
source; viz.~a very sharp cutoff at $m_\chi$ and bump-like features at
slightly smaller energies. While photons directly radiated from
charged final states give rise to a rather model-independent
contribution \cite{Birkedal:2005ep}, photons radiated from charged
virtual particles strongly depend on the details of the underlying DM
model.

As anticipated in the introduction, for what concerns the particle 
physics content, we will restrict ourselves to supersymmetric DM. 
While the Minimal Supersymmetric Standard Model (MSSM) needs more than
100 parameters for its full characterisation, in the following we will
only consider a constrained version, minimal supergravity 
(mSUGRA -- see, e.g., \cite{Chamseddine:1982jx}), where  gravity is  supposed to 
mediate SUSY breaking and the number of parameters is reduced to 4 plus a 
choice of sign after certain commonly adopted, physically well-motivated 
assumptions such as neglecting $CP$ violating or 
flavour-changing neutral current interactions and imposing the
grand unification theory (GUT) condition for the gauge couplings,
M$_1=5/3\tan^2\theta_w$M$_2\approx0.5$~M$_2$. The parameters
defining an mSUGRA model are:  universal masses for gauginos
($m_{1/2}$) and scalars ($m_0$), a common trilinear coupling term $A_0$
in the SUSY breaking part of the Lagrangian, the ratio  $\tan\beta$ of
the vacuum expectation values of the two Higgs bosons and the sign for the  
coefficient $\mu$ of the mass term in the Higgs potential. By solving
the renormalisation group equations, these parameters, defined at the
GUT scale, can be translated into masses and couplings at observable, 
i.e.~low, energies.\footnote{
  For the calculation of the low-energy features of these models 
  (i.e.~mass spectra etc.), we use version 4.01 of \ds\ \cite{Gondolo:2004sc} 
  that relies on the public code {\sffamily Isajet 7.69}
  \cite{Paige:2003mg}. One should be aware, however, that these calculations  
  are highly sensitive to how the renormalisation group equations are 
  implemented and different codes, or even different versions of the same 
  code, may give rather different results (see, e.g., Battaglia et
  al.,2001;2004b).   
  Typically, the \emph{qualitative} low-energy features of a given model can 
  still easily be reconstructed by allowing for slight shifts in the parameter 
  space (defined at high energies). From a practical point of view, this 
  situation therefore does not constitute a severe problem as one may always 
  regard a set of low-energy quantities like the mass spectrum, annihilation 
  cross section and branching ratios as a valid \emph{effective} definition of 
  the model.
}

Even if highly constrained, these models permit a rich
phenomenology. From a cosmological point of view, one can single out
five regions of the underlying parameter space that are particularly
interesting as they correspond to models with a neutralino relic density
in accordance with the value measured by WMAP:  the \emph{bulk region} 
at low $m_0$ and $m_{1/2}$, where the mass spectrum contains light
sleptons $\widetilde{\ell}$ and, as a consequence, the relic density is
mainly determined by annihilation processes
$\chi\chi\rightarrow\ell^+\ell^-$ in the early universe (through a
$t$-channel exchange of $\widetilde{\ell}$); the \emph{funnel region} at
intermediate values for $m_0$ and $m_{1/2}$, where $m_A\approx
2m_\chi$ and annihilations in the early universe are thus enhanced by
the presence of the near-resonant pseudo-scalar Higgs boson; the
hyperbolic branch or \emph{focus point region}, where $m_0 \gg
m_{1/2}$ and the neutralino becomes an almost pure Higgsino, with
large annihilation rates into gauge bosons; the \emph{stau
  coannihilation region} at large $m_{1/2}$ but small $m_0$, where
$m_{\chi}\approx m_{\widetilde{\tau}}$ and coannihilations with staus (and
usually other sleptons as well) are important in determining the relic
density; and finally the \emph{stop coannihilation region} (arising
when $A_0 \ne 0$) where $m_\chi \approx m_{\widetilde{t}}$.  

In what follows, we choose to work with a set of benchmark models,
representative of the different regions in  parameter space described
above. From an experimental point of view, the advantage of
benchmark models is that they allow a direct comparison between data
from different experiments (most of the benchmarks that we use have
already been extensively studied in other contexts) and, in general, a more 
detailed {\it per case} analysis than for, e.g., parameter scans. 
Our particular choice of benchmark models is summarised in Table~\ref{tab:bm}.  

The features of these models that are important in our context are the 
following:

\begin{itemize}
  \item $ I^{\prime} $: This model (like the following two) was
  introduced by Battaglia et al.~\cite{Battaglia:2003ab}, where also its
  phenomenology at colliders was extensively studied. It is a typical example 
  of a model in the \emph{bulk region}. While the annihilation into lepton
  pairs is strongly suppressed for neutralinos with the small
  velocities they exhibit today (unlike in the early universe),
  annihilation into $\ell^+\ell^-\gamma$, which does not suffer from
  helicity suppression \cite{Bergstrom:1989jr},  gives a considerable
  contribution due to the lightness of the sleptons.  

  \item $ J^{\prime} $: This model lies in the {\it coannihilation
  tail}. The sleptons being close to degenerate with the neutralino,
  IB from lepton final states gives even higher enhancement of the
  flux than in the previous case. 

  \item $K'$: A representative model for the {\it funnel
  region}, where the annihilation dominantly occurs through an
  $s$-channel pseudo-scalar Higgs boson. Consequently, the
  additional emission of a photon does not lift the helicity
  suppression in this case and therefore IB contributions have to be   
  subdominant.  

  \item $F^*$: Introduced in Ref.~\cite{Bringmann:2007nk} as BM4, this model
  exhibits a large neutralino mass, as typical in the {\it focus
  point} region. In this regime, the chargino is close to degenerate
  with the neutralino (in this case an almost pure Higgsino) and large IB 
  contributions result from charged
  gauge boson final states \cite{Bergstrom:2005ss}. 

  \item $J^*$: Introduced in Ref.~\cite{Bringmann:2007nk} as BM3, this is 
  another example of a neutralino in the co-annihilation region, characterised 
  by a particularly large IB contribution.
\end{itemize}

We used \ds, which in its most recent public release 5.0.1 \cite{Gondolo:2005}, 
contains a full implementation of the IB contributions focused on here, 
to compute the annihilation spectra for the benchmark models defined above. 
Line signals are also taken into account, but they turn out to be completely 
subdominant in the cases studied here (except for model $F^*$). 
The resulting spectra are plotted in Figure~\ref{fig:spectra}, both before taking 
into account the finite energy resolution of the detector and for the case 
of an energy resolution of 10\%. 
The main characteristics of these spectra are also summarised in Table~\ref{tab:bm}.

\FIGURE[htb!]{
\caption{\label{fig:spectra} 
  The particle physics factor $d\Phi^{PP}/dE$, as defined in
  Eq.~\ref{eq:particle_physics}, for the benchmarks models
  introduced in Section~\ref{subsec:phipp}. The upper panel shows the
  case of a hypothetical detector with perfect energy resolution, and
  a line width of $\epsilon\sim v\sim10^{-3}$, while the lower case
  shows the more realistic example of $\epsilon=10$\%. For comparison,
  we also show the spectrum of the Crab Nebula, taken from 
  Ref.~\cite{Albert:2007xg} with an arbitrary normalisation.}
\includegraphics[width=0.75\linewidth]{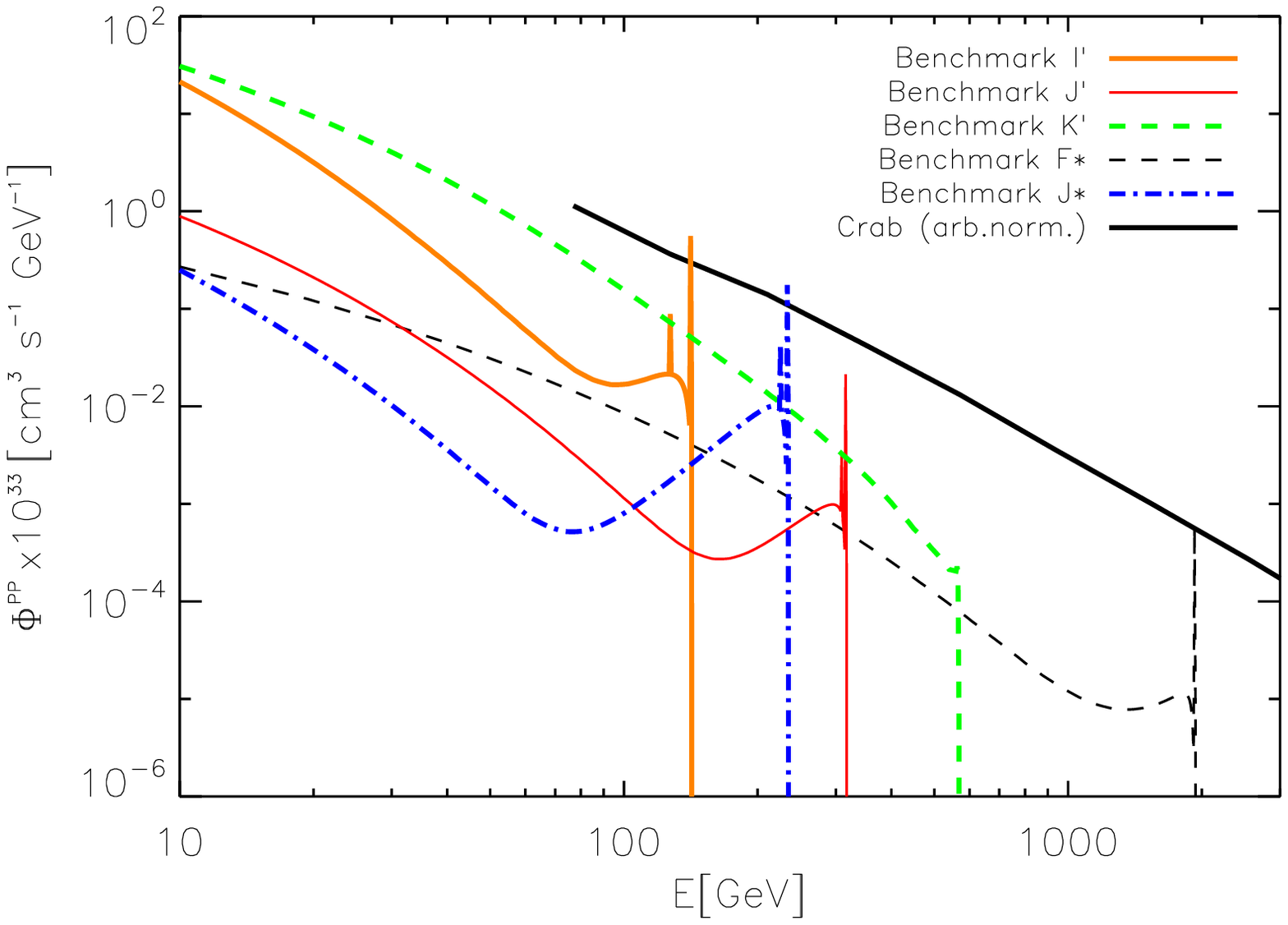}
\includegraphics[width=0.75\linewidth]{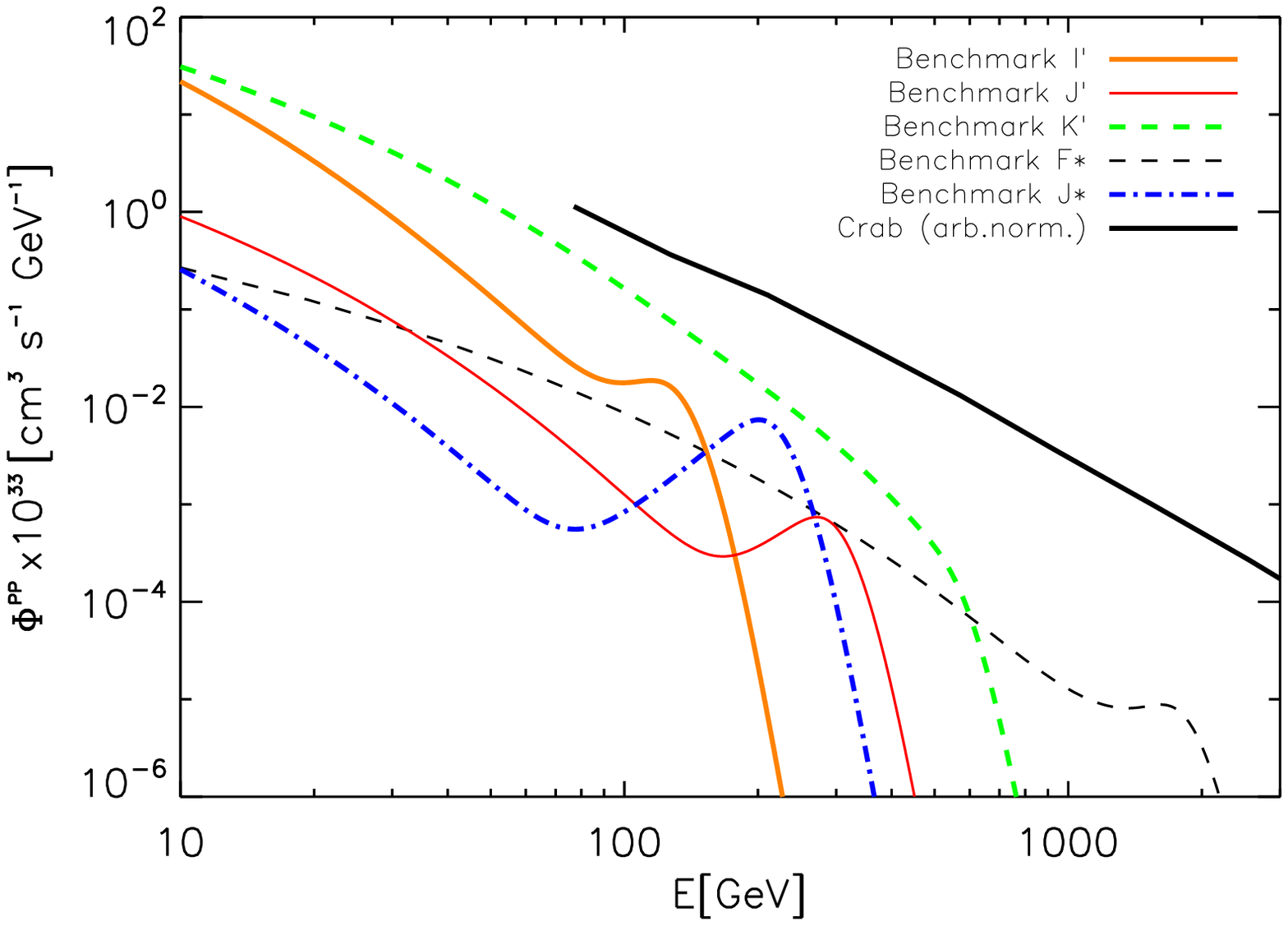}}

\section{Results and Discussion}
\label{sec:results}

\TABLE[htb!]{
\begin{tabular}{c|lc|lc}
\hline\hline
& \multicolumn{4}{c}{Draco-NFW} \\
\hline
& \multicolumn{2}{ c|}{MAGIC~II} & \multicolumn{2}{c}{CTA$_{30}$} \\
\hline
$I'$ 
& $0.75$ & ($1.9\!\cdot\!10^4,1.3\!\cdot\!10^4,\mathbf{2900}$)
& $4.7$  & ($3100,2100,\mathbf{490}$)\\
$J'$ 
& $0.10$ & ($1.4\!\cdot\!10^5,3.2\!\cdot\!10^4,\mathbf{7600}$)
& $0.52$ & ($2.8\!\cdot\!10^4,4900,\mathbf{1200}$)\\
$K'$ 
& $7.0$ & ($2000,2000,\mathbf{470}$)
& $35$  & ($410,260,\mathbf{61}$)\\
$F^*$
& $0.45$ & ($3.1\!\cdot\!10^4,1.6\!\cdot\!10^4,\mathbf{3800}$)
& $1.1$  & ($1.3\!\cdot\!10^4,2800,\mathbf{670}$) \\
$J^*$
& $0.37$ & ($3.8\!\cdot\!10^4,7400,\mathbf{1700}$)
& $0.42$ & ($3.4\!\cdot\!10^4,1200,\mathbf{290}$)\\
\hline\hline
 & \multicolumn{4}{c}{Willman~1}\\
\hline
& \multicolumn{2}{ c|}{MAGIC~II} & \multicolumn{2}{c}{CTA$_{30}$}\\ 
\hline
$I'$ 
& $1.5$  & ($9200,6200,\mathbf{150}$)
& $9.4$  & ($1500,1000,\mathbf{25}$) \\
$J'$ 
& $0.21$ & ($6.9\!\cdot\!10^4,1.6\!\cdot\!10^4,\mathbf{380}$)
& $1.1$  & ($1.4\!\cdot\!10^4,2400,\mathbf{58}$) \\
$K'$ 
& $14$  & ($990,990,\mathbf{24}$)
& $71$  & ($200,130,\mathbf{3}$) \\
$F^*$
& $0.92$ & ($1.5\!\cdot\!10^4,8100,\mathbf{190}$)
& $2.2$  & ($6500,1400,\mathbf{34}$) \\
$J^*$
& $0.76$ & ($1.9\!\cdot\!10^4,3700,\mathbf{88}$)
& $0.85$ & ($1.7\!\cdot\!10^4,610,\mathbf{15}$) \\
\hline\hline
\end{tabular}
\caption{\label{tab:flux} Expected integrated flux $\Phi(\,E>E_0$) for
our neutralino benchmark models (in units of $10^{-15}$~ph
cm$^{-2}$ s$^{-1}$), where we used the experimental parameters
listed in Table~\ref{tab:IACT}. In parentheses, we state the
increase in the signal that would be needed for a $5\sigma$
detection as (B1,$B2$,{\bf B3}). Here, B1 is the often cited
increase that is needed when simply comparing the sensitivity
and annihilation fluxes above the telescope energy threshold
$E_0$. $B2$ is more realistic in that it gives the corresponding
quantity \emph{above a certain energy $E^*$, depending on the
benchmark, where the integrated flux to sensitivity ratio is
greatest} and {\bf B3} is the same as $B2$, yet for the most
favourable halo profile consistent with current observations
(still not taking into account the effect of substructures,
however). See text for further details.} 
}


Combining the astrophysical factor of Table~\ref{tab:jpsi} and the particle 
physics factor from Table~\ref{tab:bm}, we can finally make predictions about the 
expected  gamma ray flux above the telescope energy threshold 
$E_0$. A summary of the results is reported in Table~\ref{tab:flux}, where  
we also quote the increase in the overall flux normalization that would be necessary to meet the 
required sensitivity for a $5\sigma$ detection (referred to as B1 in the 
table). While it is customary to quote sensitivities and actual fluxes above 
$E_0$ in this kind of analysis, we recall that DM annihilation spectra are 
rather hard, in particular when taking into account possible spectral 
features at photon energies close to the spectral cutoff at the mass of the DM
particle. On the other hand, the sensitivity of IACTs is considerably
better at energies somewhat larger than the telescope energy
threshold. We therefore take the projected sensitivities for the
integrated flux above some energy $E^*>E_0$, using the sensitivity
curves as provided by Bernl\"oher et al.~\cite{Bernloher:2007} for CTA and
Carmona et al.~\cite{Carmona:2007rs} for MAGIC~II and, by 
comparing those to the annihilation spectra, compute the
\emph{minimal} increase in the normalisation that is required to
see at least part of the DM annihilation spectrum above $E^*$. This 
is referred to as the quantity $B2$ in Table~\ref{tab:flux}; finally, we also 
state as {\bf B3} the corresponding value for the most favourable 
\emph{smooth} halo profile that is consistent with the observational data (i.e.~here we take the upper limit on $\tilde J$ as 
discussed in Section~\ref{subsec:astro}).

So far, we have only discussed smooth DM distributions. On the other hand, it 
is well known from both theory \cite{Green:2005fa} and numerical $N$-body 
simulations \cite{Diemand:2005vz} that cold DM is expected to cluster and 
thereby to form substructures with masses all the way down to the small--scale 
cutoff in the spectrum of matter density fluctuations, which can be 
determined to a great accuracy from the underlying DM model 
\cite{Bringmann:2006mu}; if surviving until today, such inhomogeneities in the 
DM distribution would greatly enhance the DM annihilation rate
\cite{Bergstrom:1998jj}. For the case of typical dSphs, this could 
result in  an additional boost of the signal by a factor of 10-100 
\cite{Strigari:2006rd}. Another considerable boost in the
annihilation flux  could also result from the existence of a
hypothetical black hole at the center of the dwarfs
\cite{Colafrancesco:2006he}. In the most optimistic astrophysical 
configuration, the required increase stated as $\mathbf{B3}$ in
Table~\ref{tab:flux}, would thus further be \emph{reduced} by up to
two orders of magnitude. 

Let us now discuss some details of the results:

\begin{itemize}
\item {\it Sources.} For Draco, the model-dependent fluxes for the Burkert 
and NFW profiles are very similar, and therefore we presented only the latter 
in Table~\ref{tab:flux}. For the astrophysical benchmark profiles introduced 
in Section~\ref{subsec:astro}, detectional prospects  for Draco and Willman~1
only differ by a factor of around 2, and are obviously not very encouraging. 
When considering the most optimistic astrophysical configurations, adopting the highest observationally allowed value for $\tilde J$, 
things change considerably and Willman~1 becomes an interesting and indeed very 
promising target for DM searches. Allowing for an additional, in fact 
well-motivated, boost due to the presence of DM substructures in the dwarfs, 
this may give at least CTA the chance to see also Draco in some cases.

\item {\it Telescopes.}  Depending on the DM model, the ability of CTA to
detect gamma rays from DM annihilation is a factor of $6-8$ better than for  
MAGIC~II. Focusing on Willman~1, and assuming very favourable astrophysical 
conditions, CTA would in principle be able to see \emph{all} the benchmark 
models considered here, while MAGIC~II should be able to see at least some 
of them.
We recall that the flux enhancements needed for a $5\sigma$ detection, as states in Table~\ref{tab:flux}, are calculated with respect 
to an observation time of $t_{obs}=50$~hrs and scale like $t_{obs}^{-1/2}$. 
For prolonged observation times, one could thus win a factor of a few for 
both telescopes. Furthermore, as the CTA parameters are still quite 
preliminary, an additional factor of 2 in the sensitivity of the operating 
instrument seems quite feasible.

\item {\it Benchmark models.} The best prospects for detection are found for 
the neutralino in the funnel region (model $K'$), the reason simply being 
a rather large annihilation rate.
The second-best prospects are found for model $J^*$ in the coannihilation 
region. Recalling from Table~\ref{tab:bm} that $J^*$ is actually the model 
with the \emph{smallest} annihilation rate, this may come as some surprise 
and nicely illustrates the importance of including IB contributions when 
estimating the flux from DM annihilation.
The model $F^*$ is yet another example with rather pronounced IB 
contributions; a mass of almost $2~$TeV, however, efficiently suppresses the 
annihilation flux (in this case, the required boost actually depends 
significantly on the details of the --- so far not sufficiently well
known  --- integrated sensitivity of CTA for TeV photons  and could
thus eventually be significantly improved).
\end{itemize}

When compared to previous work, we thus arrive at considerably more optimistic 
conclusions than what has been reached  by Sanchez--Conde et al.~\cite{SanchezConde:2007te} 
for the analysis of present-day gamma ray telescopes -- not the least due to 
our fully taking into account all the contributions to the expected 
annihilation spectrum. 
On the other hand, we find the conclusions of Strigari et al.~\cite{Strigari:2007at}
overly optimistic, a fact that we trace back to the very large particle flux 
factor of 
$\Phi^{PP}=5\times 10^{-29}\mathrm{cm}^3\mathrm{s}^{-1}\mathrm{GeV}^{-2}$ 
that the authors assumed as a fiducial value (this should be compared to 
Table~\ref{tab:bm} and the corresponding values for our benchmark models, 
which represent typical neutralino DM candidates). While it 
may indeed be possible to find DM models with higher gamma ray yields than 
considered here, we recall that there exist rather tight general bounds on 
the allowed annihilation cross section and the number of high-energy photons 
that are produced \cite{Mack:2008wu}. Finally, 
we would like to remark that it is certainly promising to combine the DM 
searches in gamma rays, as described here, with observations of the same 
dSphs at other wavelengths 
(see also \cite{Colafrancesco:2006he,Jeltema:2008ax}).


\section{Conclusions}
\label{sec:conclusions}
In this article, we have computed the prospects of detecting gamma rays from 
annihilating DM in two nearby dwarf galaxies, Draco and Willman~1, for the 
upcoming ground--based Imaging Atmospheric Cherenkov Telescopes MAGIC~II and 
the CTA telescope array (the latter still being in the early design phase). We 
have focused our analysis on a set of five benchmark models, representatives 
for the parameter space of neutralino DM in the mSUGRA framework, and paid
special attention to describing those telescope features that are 
most relevant in this context. For the first time in this kind of analysis, 
we have fully taken into account the contributions from radiative corrections 
that were recently reported by Bringmann et al.~\cite{Bringmann:2007nk}.
As it turned out, in fact, taking realistic DM spectra has an important impact 
on the analysis and, although common practice, \emph{it can be a rather bad 
approximation to simply assume a featureless DM spectrum like from $b\bar b$ 
fragmentation and/or to only focus on the total flux above a given energy 
threshold $E_0$} in these kind of studies. The basic underlying reason for 
this is that \emph{realistic DM annihilation spectra show a harder energy 
dependence than the sensitivity of IACTs}.
Once detected, clear spectral features would, of course, have the additional 
advantage of providing a rather fool-proof way of discriminating DM spectra 
against astrophysical background sources -- which is even more important in 
view of the still rather large astrophysical uncertainties involved.

Although these effects do provide a considerable enhancement of the 
detectional prospects, the expected flux from dSphs remains at a level that, for 
conservative scenarios, will be challenging to detect with the next generation 
of IACTs. This, rather than the angular resolution of these instruments, is 
the reason why the potential of IACTs to
discriminate between different DM profiles in  dSphs is limited even in 
the case of the detection of an annihilation signal; the  eventual 
disentanglement between cored and cuspy profiles is probably more promising to 
perform at other wavelengths. 

On the other hand, if one adopts the most optimistic astrophysical 
configurations that are compatible with current observational data of 
Willman~1, i.e. a favourable DM profile and an $\mathcal{O}(10-100)$ flux 
enhancement due to the existence of substructures, \emph{all} of our 
benchmark models approach the reach of at least the CTA which, for the models 
studied here, is a factor of $6-8$ more sensitive to the annihilation signal 
than MAGIC~II (this is, of course, independent of the source). The most 
promising case of our analysis turns out to be a neutralino in the funnel 
region, characterised by no sizeable IB contributions to its spectrum but a 
rather large annihilation rate; the second best case is a neutralino from the 
coannihilation region, making up for its small annihilation rate with 
enormously large radiative corrections.

Having demonstrated that the prospects of indirect DM detection through gamma 
rays do depend on the details of the annihilation spectrum, and thus the 
underlying particle nature, it would be interesting to perform similar 
analyses also for other targets of potential DM annihilation. 
Another further direction of extending the present analysis would be to 
perform a full scan over the parameter space of viable 
models. Finally, we have stressed that the very concept of sensitivity of an 
IACT depends on the spectrum that is observed; in the context of DM searches, 
this is particularly important as DM annihilation spectra can significantly 
deviate from the usually assumed Crab-like spectrum.
While we have provided a first estimate of how to proceed in such a case, 
it would be warranting to perform a dedicated analysis, using the full power 
of state-of-the-art MC tools, in order to accurately determine the importance 
of this effect.

Finally, we would like to mention that even in the case of negative detection,
IACTs could in principle put interesting upper limits on the flux which in 
turn would translate into constraints on the combined space of astrophysical 
and particle physics parameters. Though much smaller than for other sources 
like, e.g., the galactic center, the main uncertainty in this case lies in 
the overall scale of the flux as determined by the details of the DM 
distribution. This, unfortunately, will therefore greatly obstacle any 
stringent constraint from null searches on the particle physics nature of DM
for quite some time ahead.

To conclude, nearby dwarf galaxies -- and in particular Willman~1 -- are very 
interesting and promising targets for DM searches with the next generation of 
IACTs. An excellent performance of these experiments, in particular in terms 
of the sensitivity at energies slightly below the DM particle mass, will be 
paramount in such searches. In fact, given the low level of fluxes involved, 
a factor of 2 in sensitivity might decide whether a signal will be seen or 
not. Complementary to such demanding requirements on the experiments, the 
above discussion should also have made clear that  it will be very important 
to collect more astrophysical data and to improve the theoretical 
understanding of how DM is distributed in order to reduce the still 
unpleasantly large astrophysical uncertainties involved.

\acknowledgments
We would like to thank the MAGIC collaboration and in particular G. Bertone, 
E. Carmona, M. Mariotti, M. Persic, L. Pieri for useful suggestions and A. 
Biland, D. Mazin and the anonymous referee for helpful discussions and
comments on the manuscript.\\

\begin{footnotesize}
\emph{Note added:} After the completion of this work, Segue~1 has been
presented as yet another ultra-faint Milky Way dwarf satellite galaxy,
with an expected DM induced gamma-ray flux almost twice that of Willman~1
\cite{Geha:2008zr}. Our conclusions can be easily applied to this interesting
newly discovered target by simply scaling the boost factors reported in
Table~\ref{tab:flux} accordingly. We also mention that the MAGIC collaboration
recently reported an upper limit on the observation of Willman~1 of the order
of $ 10^{-12} \mbox{ph cm}^{-2} \mbox{s}^{-1} $ above 100 GeV using the
analysis method described in \cite{Aliu:2008ny}.

\end{footnotesize}

\bibliographystyle{JHEP} 
\bibliography{biblio} 
\end{document}